\pdfoutput=1
\documentclass[man,12pt,nolmodern,floatsintext]{apa7}
\usepackage{amsmath,amsthm,amssymb}
\usepackage{graphicx,float}
\usepackage[nodisplayskipstretch]{setspace}
\usepackage{enumitem}
\usepackage[american]{babel}
\usepackage{color}
\usepackage{verbatimbox}
\usepackage{caption}
\usepackage{mathtools}
\usepackage{multirow}
\usepackage{array}
\usepackage{xparse}
\usepackage{csquotes}
\usepackage{fancyhdr}
\usepackage{float}
\usepackage{url}
\usepackage[style=apa,backend=biber]{biblatex}
\usepackage{bm}
\usepackage{makecell}
\usepackage{booktabs}
\usepackage[normalem]{ulem}
\usepackage{hyperref}
\hypersetup{hidelinks}
\restylefloat{table}
\restylefloat{figure}

\usepackage{pdfpages}

\definecolor{blue1}{RGB}{40,80,156}
\definecolor{red1}{RGB}{228,55,55}

\newcommand{\ex}{\mathbb{E}}
\newcommand{\pr}{\mathbb{P}}
\newcommand{\xx}{\mathbf{x}}
\newcommand{\yy}{\mathbf{y}}
\newcommand{\YY}{\mathbf{Y}}

\def\bY{\mathbf{Y}}

\def\var{\mathrm{Var}}

\def\t{^\top}
\def\xx{\mathbf{x}}
\def\yy{\mathbf{y}}

\def\nnu{\boldsymbol{\nu}}
\def\eeta{\boldsymbol{\eta}}

\def\cII{\boldsymbol{\mathcal{I}}}

\DeclareMathAlphabet{\mathbf}{OT1}{cmr}{b}{n}

\long\def\mk#1{\bgroup\color{black}#1\egroup}
\long\def\mkr#1{\bgroup\color{black}#1\egroup}

\title{Asymptotic Standard Errors for Reliability Coefficients in Item Response Theory}
\shorttitle{IRT Reliability}
\authorsnames{Youjin Sung, Yang Liu}
\authorsaffiliations{Department of Human Development and Quantitative Methodology, University of Maryland}

\authornote{Correspondence should be made to Youjin Sung at 1232 Benjamin Bldg, 3942 Campus Dr, University of Maryland, College Park, MD, 20742. Email: yjsung@umd.edu.}

\abstract{In a recent review, Liu, Pek, \& Maydeu-Olivares (2025b) classified reliability coefficients into two types: classical test theory (CTT) reliability and proportional reduction in mean squared error (PRMSE). This article focuses on quantifying the sampling variability of these coefficients under item response theory (IRT) models. \mkr{While some existing standard error (SE) formulas are accurate when variability arises only from item parameter estimation, the reliability estimators considered in our work involve additional variability from substituting population moments with sample moments.} We propose a general strategy to derive SEs that incorporates both sources of sampling error simultaneously, enabling the estimation of model-based reliability coefficients and their SEs \mkr{in such settings}. We then apply our general theory to derive SEs for two specific estimators under the graded response model: (1) CTT reliability for the expected \textit{a posteriori} score of the latent variable and (2) PRMSE for the latent variable. Simulation results show that the derived SEs accurately capture the sampling variability across various test lengths in moderate to large samples. We conclude with an empirical illustration and directions for future research.}
\keywords{reliability, item response theory, asymptotic standard errors}

\begin{document}
                \maketitle
\doublespacing

\section{Introduction}

Reliability is an overall index of measurement precision \parencite{standards}. Given a latent variable (LV) measurement model, a reliability coefficient quantifies how well observed scores\footnote{\mkr{Throughout the paper, the term ``observed scores'' refers to scores that are functions of response variables, encompassing not only summed scores but also factor scores that depend on model parameters. This interpretation aligns with Liu et al. (2025a, 2025b).}}, which are functions of response variables, align with latent scores, which are functions of LVs that reflect constructs  of interest \parencite{liu2025general}. Recently, \textcite{liu2024} introduced a regression-based framework of measurement precision, which defines reliability as coefficients of determination associated with regressions. Within this framework, classical test theory (CTT) reliability corresponds to the coefficient of determination when regressing an observed score onto all LVs in the measurement model. Meanwhile, proportional reduction in mean squared error \parencite[PRMSE;][]{haberman2010reporting}, another popular index of measurement precision in item response theory \parencite[IRT;][]{thissen2009}, corresponds to the coefficient of determination when regressing a latent score onto all response variables.\footnote{\textcite{liu2024} reserved the term ``reliability'' exclusively for CTT reliability. To be more consistent with the IRT literature \parencite[e.g.][]{haberman2010reporting, liu2025general}, we treat both CTT reliability and PRMSE as reliability coefficients in this article.} 

While \citeauthor{liu2024}'s (\citeyear{liu2024}) regression formulation of CTT reliability and PRMSE provides a conceptual framework for understanding reliability, in practice, reliability coefficients are estimated from sample data and thus are subject to sampling variability. This inherent uncertainty should be quantified and communicated through standard errors (SEs) and confidence intervals \parencite[CIs; e.g.,][]{fan2001confidence}; however, existing literature on this topic is sparse. The primary goal of this paper is to address this gap by analytically deriving asymptotic SEs for CTT reliability and PRMSE, assuming an IRT model as the underlying LV measurement model.

So far, only a few studies have focused on computing SEs or CIs for reliability coefficients within the IRT framework. \textcite{andersson2018} derived asymptotic SEs for the so-called marginal reliability \parencite{cheng2012comparison} and test reliability \parencite{kim2010estimation} using the Delta method, both of which can be viewed as forms of CTT reliability for specific observed scores, as discussed in later sections. Alternatively, \textcite{yang2012} obtained SEs for the marginal reliability based on multiple imputation. \mkr{These approaches provide valid uncertainty quantification when the reliability coefficient can be expressed solely as a function of item parameters; namely, when sampling variability arises only from item parameter estimation.

However, the formulas of CTT reliability and PRMSE involve population moments (e.g., variances) implied by the measurement model (Liu et al., 2025a, 2025b). Computing these population moments, given estimated item parameters, is feasible for relatively short tests, but quickly becomes computationally infeasible as test length increases because the total number of possible response patterns grows exponentially.} Although certain population variances can be approximated via increasing-test-length asymptotics (e.g., for marginal reliability), such approximations are typically tied to specific types of scores (e.g., maximum likelihood scores) and are useful only under limited circumstances (e.g., when the test is very long).  
Consequently, when reliability is viewed more generally through the regression framework and the test involves many items, a preferable alternative is to substitute population moments with sample estimates \parencite[e.g., empirical reliability;][]{chalmers2012mirt}.

To provide valid uncertainty quantification in such scenarios, this study presents a general SE derivation for reliability coefficients that are subject to two sources of sampling variability: item parameter estimation and the use of sample moments. In theory, our method is applicable to CTT reliability for any observed score and PRMSE for any latent score, under any fully specified parametric IRT model. We provide full derivations for two specific reliability coefficients as examples in the present work: (1) CTT reliability for the expected \textit{a posteriori} (EAP) score of the LV and (2) PRMSE for the LV. The derivations are provided under the unidimensional graded response model \parencite[GRM;][]{samejima1969estimation} for polytomous items, which encompasses the two-parameter logistic (2PL) model \parencite{birnbaum1968some} as a special case for dichotomous items. While not explicitly addressed here, the proposed derivation readily extends to other observed or latent scores under other types of parametric IRT models.

The remainder of the paper is structured as follows. 
We begin by introducing IRT and provide a brief review of the regression framework of reliability. We then summarize how existing IRT reliability coefficients can be classified under the regression framework. After reviewing the literature of SE estimation for reliability coefficients, we present the general theoretical framework for deriving SEs and apply it to the two example reliability coefficients under the GRM. The finite sample performance of the derived SEs is then evaluated in a simulation study for both the 2PL model and the GRM, followed by an empirical illustration. Finally, we conclude with a summary and a discussion of the limitations and potential extensions of this study.

\section{IRT Reliability from a Regression Framework}

\subsection{Item Response Theory}

Let $\Theta_i$ denote a unidimensional LV for person $i$, which is assumed to follow a standard normal distribution. 
Let $Y_{ij}$ denote a random response variable for person $i$ on item $j$ and $\bY_i=(Y_{i1}, \dots, Y_{im})^{\t}$ be a collection of $m$ response variables from individual $i$. Let the corresponding lowercase letters $\theta_i, y_{ij},$ and $\yy_i$ indicate the realizations of $\Theta_i, Y_{ij},$ and $\bY_i$, respectively.
Conditioned on $\Theta_i=\theta_i$, it is assumed that $Y_{ij}$, $j=1, \dots, m$, are independent (i.e., local independence; \cite{mcdonald1981dimensionality}).

Let item $j$ have $K$ ordered response categories scored $0, \dots, K-1$. The conditional probability of endorsing category $k$ (i.e., $Y_{ij}=k\in\{0, 1,  \dots,K-1\}$) given $\theta_i$ is parameterized by a graded response model \parencite[GRM;][]{samejima1969estimation}:
\begin{equation}
f_j(k | \theta_i; \nnu)
= f_j^{*}(k | \theta_i; \nnu)
 - f_j^{*}(k+1 | \theta_i; \nnu),
 \label{eq:grmfinal}
\end{equation}
where
\begin{equation}
    f_j(k|\theta_i; \nnu)=\pr\{Y_{ij}=k|\theta_i; \nnu\}
\end{equation}
and 
\begin{equation}
f_j^{*}(k | \theta_i; \nnu)
= \pr\{Y_{ij} \ge k|\theta_i; \nnu\} 
=
\begin{cases}
1, & k=0, \\[6pt]
\dfrac{1}{1+\exp\!\left[-(a_j\theta_i + c_{jk})\right]}, & k= 1, \dots, K-1, \\[10pt]
0, & k=K.
\end{cases}
\label{eq:grm_2pl}
\end{equation}
\mkr{Equation \ref{eq:grm_2pl} and the subsequent equations use the slope-intercept parameterization}, where $a_j$ is the slope parameter for item $j$, and $c_{jk}$ is the $k$th intercept parameter for item $j$ that is related to the $k$th item difficulty parameter $b_{jk}=-c_{jk}/a_j$. The intercept parameters satisfy the ordering constraint
\begin{equation}
c_{j1} > \cdots > c_{j,K-1},
\end{equation}
which ensures that 
\begin{equation}
f_j^{*}(k | \theta_i; \nnu) > f_j^{*}(k+1 | \theta_i; \nnu), \quad k=0, \dots, K-1, \quad \text{for all } \theta_i.
\end{equation}
The vector $\nnu$ collects all item parameters (i.e., $m$ slopes and $(K-1)m$ intercepts) into a $Km \times 1$ vector. When $K=2$, the GRM reduces to the two-parameter logistic (2PL) model \parencite{birnbaum1968some}.

Under this model, the marginal likelihood of person $i$'s responses $\yy_i=(y_{i1},\dots,y_{im})^{\t}$ is expressed by 
\begin{equation}
    f(\yy_i; \nnu)=\int f(\yy_i|\theta_i;\nnu)\phi(\theta_i)d\theta_i,
    \label{eq:marginal}
\end{equation}
in which
\begin{equation}
    f(\yy_i|\theta_i;\nnu)=\prod_{j=1}^{m} f_j(y_{ij}|\theta_i;\nnu)
    \label{eq:cond}
\end{equation}
is the conditional likelihood of $\yy_i$ given $\theta_i$, and 
$\phi$ is the density of $\mathcal{N}(0, 1)$.

Given a sample of $n$ independent and identically distributed (i.i.d.) random vectors of responses $\YY_1,\dots,\YY_n$, we express the sample log-likelihood as
\begin{equation}
   \hat\ell(\nnu) = \frac{1}{n}\sum_{i=1}^n \log f(\YY_i; \nnu).
   \label{eq:LL}
\end{equation}
In IRT models, $\nnu$ is often estimated by maximum likelihood (ML), in which the estimates for $\nnu$ are found by solving the following estimating equation 
\begin{equation}
    \nabla_{\nnu}\hat\ell(\nnu)  = \mathbf{0}.
    \label{eq:est_eq}
\end{equation}
In Equation \ref{eq:est_eq}, $\nabla_{\nnu}\hat\ell(\nnu)$ denotes a $Km \times 1$ vector of partial derivatives of $\hat\ell(\nnu)$ with respect to $\nnu$. Given the negative definiteness of the Hessian matrix, the solution to Equation \ref{eq:est_eq}, denoted by $\hat{\nnu}$, is the ML estimator of $\nnu$, a local maximizer of the log-likelihood function under suitable regularity conditions. Given correct model specification, $\hat{\nnu}$ satisfies 
\begin{equation}
  \sqrt{n}(\hat{\nnu}-\nnu_0) = \cII^{-1}(\nnu_0)\sqrt{n}\nabla_{\nnu_0}\hat\ell(\nnu_0) + o_p(1) \overset{d}{\to} \mathcal{N}\left(\mathbf{0},\cII^{-1}(\nnu_0) \right)
    \label{eq:info}
\end{equation}
as $n \to \infty$, where $\nnu_0$ denotes true parameters and $\cII(\nnu_0)=\ex [\nabla_{\nnu_0} \log f(\bY_i; \nnu_0) \nabla_{\nnu_0} \log f(\bY_i; \nnu_0)^{\t}]$ denotes the $Km \times Km$ (per-observation) Fisher information matrix. 

The LV $\Theta_i$ is often predicted based on its posterior density given the observed responses $\yy_i$:
\begin{equation}
    f(\theta_i|\yy_i;\nnu) = \frac{f(\yy_i|\theta_i;\nnu)\phi(\theta_i)}{f(\yy_i;\nnu)}.
    \label{eq:post}
\end{equation}
A commonly used example is the expected \textit{a posteriori} (EAP) score, which is the mean of Equation \ref{eq:post} and expressed as follows:
\begin{equation}
    \ex(\Theta_i|\yy_i; \nnu) = \frac{\int \theta_i f(\yy_i|\theta_i;\nnu)\phi(\theta_i)d\theta_i}{f(\yy_i;\nnu)}.
     \label{eq:eap}
\end{equation}
Note that the integrations involved in Equations \ref{eq:marginal} and \ref{eq:eap} are often approximated by numerical quadrature \parencite{darrell1970fitting} as closed-form expressions do not exist. 

\subsection{Reliability Coefficients from a Regression Perspective}

Adapting the terms and notations of \textcite{liu2024}, we define an \textit{observed score} $s(\yy_i)$ as a function of response variables $\textbf{y}_i$ and a \textit{latent score} $\xi(\theta_i)$ as a function of the LV $\theta_i$. Inspired by \textcite{mcdonald2011measuring}, \textcite{liu2024} proposed a regression-based framework of reliability, in which reliability coefficients are defined as coefficients of determination associated with regressions. Specifically, the measurement decomposition of an observed score $s(\yy_i)$ concerns regressing $s(\yy_i)$ on the LV $\theta_i$ (or equivalently, on the true score underlying $s(\yy_i)$), and the prediction decomposition of a latent score
 $\xi(\theta_i)$ concerns regressing $\xi(\theta_i)$ on all response variables $\yy_i$ (or equivalently, on the EAP score of $\xi(\theta_i)$). The measurement decomposition results in reliability defined under classical test theory (CTT), while the prediction decomposition results in proportional reduction in mean squared error \parencite[PRMSE;][]{haberman2010reporting}. In the following subsections, we provide a brief summary of these two formulations under the IRT model presented in the previous section. A rigorous justification for these regression formulations can be found in the Supplementary Materials of \textcite{liu2024}. 

\paragraph{Measurement Decomposition.}

The measurement decomposition of an observed score $s(\yy_i)$ is expressed as 
\begin{equation}
    s(\yy_i)=\ex[s(\bY_i)|\theta_i] + \varepsilon_i,
    \label{eq:measdecomp}
\end{equation}
which is also known as the true score formula \parencite[e.g.,][]{raykov2011introduction}. As elaborated in \textcite{liu2024}, Equation \ref{eq:measdecomp} can be viewed as a nonlinear regression of $s(\bY_i)$ onto the LV $\Theta_i$, or equivalently, as a linear regression of $s(\bY_i)$ onto its true score $\ex[s(\bY_i)|\theta_i]$ with a zero intercept and unit slope. 

In either case, the corresponding coefficient of determination is given by the ratio of the true score variance to the observed score variance, aligning with the well-known definition of CTT reliability:
\begin{equation}
    \text{Rel}(s)=\frac{\var(\ex[s(\bY_i)|\Theta_i])}{\var [s(\bY_i)]}.
    \label{eq:cttrel}
\end{equation}
It is highlighted by the notation that CTT reliability is a property of the observed score $s$. It measures how well the chosen observed score reflects the underlying LV that is assumed to have generated it.

\paragraph{Prediction Decomposition.}

The prediction decomposition of a latent score $\xi(\theta_i)$ is expressed by 
\begin{equation}
   \xi(\theta_i)=\ex[\xi(\Theta_i)|\yy_i] + \delta_i,
   \label{eq:preddecomp}
\end{equation}
in which  $\ex[\xi(\Theta_i)|\yy_i]$ is the EAP score of $\xi(\Theta_i)$. In the special case of primary interest, $\xi(\Theta_i)=\Theta_i$, its EAP score is given by Equation \ref{eq:eap}. The error term $\delta_i$ represents the remaining uncertainty in the latent score after being predicted from the observed responses $\yy_i$, capturing the prediction error.   
Analogous to the measurement decomposition, Equation \ref{eq:preddecomp} can be viewed as a nonlinear regression of $\xi(\Theta_i)$ on $\bY_i$, or equivalently, as a unit-weight linear regression of $\xi(\Theta_i)$ on its EAP score. 

In either case, the associated coefficient of determination is given by the ratio of the EAP score variance to the latent score variance, which is identical to PRMSE \parencite{haberman2010reporting}:
\begin{equation}
    \text{PRMSE}(\xi)
    =\frac{\var(\ex[\xi(\Theta_i)|\bY_i])}{\var[\xi(\Theta_i)]}.
    \label{eq:prmse}
\end{equation}
As emphasized by the notation, PRMSE is inherently a property of the latent score $\xi$ and quantifies how much uncertainty in $\xi$ is reduced when making the optimal prediction using the available data $\yy_i$.

\paragraph{Connections to Existing IRT Reliability Coefficients.}

For unidimensional IRT models, CTT reliability and PRMSE have been referred to under various names in the literature. First, CTT reliability is equivalent to ``parallel-forms reliability'' \parencite{kim2012note}, defined as the correlation between LV estimates (i.e., observed scores in the regression framework) from two parallel forms of a test.\footnote{Connections between the formulation in \textcite{kim2012note} and the regression-based formulation were discussed in the Supplementary Materials of \textcite{liu2024}.} It is also equivalent to ``marginal reliability'' defined by \textcite{green1984technical}, as noted in \textcite{kim2012note}. 
Distinct terms have also been used for CTT reliability, or its approximations, when it is applied to specific types of observed scores. For example, CTT reliability for the summed score has been referred to as ``test reliability'' \parencite{kim2010estimation}. As another example, ``marginal reliability'' has also been used to describe an approximation to CTT reliability for the ML score of the LV \parencite[e.g.,][]{cheng2012comparison, andersson2018, yang2012}. This index is only an approximation because the inverse test information replaces the variance of the ML score in the formula, and the two coincide only in tests of infinite lengths.

Another type of reliability discussed in \textcite{kim2012note} is the ``squared-correlation reliability'', defined as the squared correlation between the LV and its estimate. When the EAP score serves as the LV estimate, the ``squared-correlation reliability'' coincides with the PRMSE of the LV \parencite{liu2024}. In practice, the moments in the PRMSE formula are often replaced with sample moments, as the exact calculation of the population version becomes increasingly burdensome as test length grows. The \texttt{mirt} package \parencite{chalmers2012mirt} in R \parencite{R} provides a sample moment–based estimate of the measure under the label ``empirical reliability.'' When ``empirical reliability'' is reported for other types of LV estimates, however, it serves only as an approximation to the exact form of PRMSE, which applies exclusively when the optimal predictor of the LV (i.e., the EAP score) is used in the calculation of the squared correlation.\footnote{The EAP score is optimal in that it minimizes the mean square error (MSE) in the regression defined in Equation \ref{eq:preddecomp}. Using other types of observed scores as the regressor means inefficient use of the available information, which would result in suboptimal prediction.} 

\section{Asymptotic Standard Errors of IRT Reliability Coefficients}

The derivation of SEs for IRT reliability coefficients has been very limited in scope. Based on standard large-sample theory, \textcite{andersson2018} derived asymptotic SEs for ``marginal reliability'' \parencite[][]{cheng2012comparison} (i.e., an approximation to CTT reliability for the ML score) and ``test reliability'' \parencite{kim2010estimation} (i.e., CTT reliability for the summed score). In both cases, reliability coefficients can be expressed as transformations of item parameters, and the Delta method is employed to capture sampling variability arising from item parameter estimation. This approach still applies to computing ``marginal reliability'' in long tests, as the coefficient depends on test information rather than the exact population variance of ML scores. In addition, the method remains applicable to ``test reliability'' because the number of possible summed scores increases linearly rather than exponentially with test length. However, the exact forms of CTT reliability and PRMSE typically require population-level moment calculations over all possible response patterns, making the use of sample moments unavoidable for long tests. This constraint limits the applicability of the existing approach to other reliability coefficients not addressed in the original work.\footnote{SEs of reliability coefficients for the ML and summed scores defined based on multiple-group IRT models have also been derived by \textcite{andersson2022reliability}. However, the derivation is still based on the Delta method. Alternatively, \textcite{yang2012} estimated SEs for ``marginal reliability'' by simulation, which is subject to the same limitation.}

To address this limitation, we focus on deriving SEs for reliability estimators in which population moments are replaced with sample counterparts. In this setting, sampling variability arises from both item parameter estimation and the use of sample moments, requiring more elaborate derivations. In the following subsections, we begin by
presenting a general asymptotic normality result for a family of parameterized sample
statistics, which applies to CTT reliability and PRMSE for any type of observed or latent score. 
This general theory is then applied to two specific examples under the GRM: (1) CTT reliability for the EAP score of the LV (i.e., $ s(\yy_i; \nnu) = \ex(\Theta_i|\yy_i;\nnu)$ in Equation \ref{eq:measdecomp}) and (2) PRMSE for the LV (i.e., $ \xi(\Theta_i)=\Theta_i$ in Equation \ref{eq:preddecomp}). The derivation for PRMSE is presented first due to its relative simplicity.
The dependency on the model parameters $\nnu$ is now explicitly displayed in each of the formulas.

\subsection{General Theory}

For each person $i$, consider a $\ell \times 1$ random vector
\begin{equation}
    \mathbf{H}(\bY_i; \nnu)=(H_1(\bY_i; \nnu), H_2(\bY_i; \nnu), \dots, H_{\ell}(\bY_i; \nnu))^\top,
\label{eq:g.H}
\end{equation}
in which each component $H_s(\bY_i; \nnu), \,s=1, \dots, \ell$, is a real-valued function that is allowed to depend on both the response variables $\bY_i$ and the model parameters ${\nnu}$. Additionally, we assume that $H_s(\yy_i; \nnu)$ is differentiable in $\nnu$ for every response pattern $\yy_i$. The population expectation of $H_s(\YY_i; \nnu)$ is given by 
\begin{equation}
    \eta_s(\nnu)=\ex[H_s(\bY_i; \nnu)]=\sum_{\yy_i} H_s( \yy_i; \nnu) f(\yy_i; \nnu),
\label{eq:eta_single}
\end{equation}
in which the summation is taken over all possible response patterns. This quantity can be estimated from sample data using the sample average:
\begin{equation}
    \hat{\eta}_s(\nnu)=\frac{1}{n}\sum_{i=1}^n H_s(\bY_i; \nnu).
\label{eq:etahat_single}
\end{equation}
Extending Equations \ref{eq:eta_single} and \ref{eq:etahat_single} to the vector form, let $\eeta(\nnu)$ and $\hat\eeta(\nnu)$ denote the population expectation and the sample average of $\mathbf{H}(\bY_i; \nnu)$, respectively:
\begin{align}
\eeta(\nnu) &= (\eta_1(\nnu), \dots, \eta_{\ell}(\nnu))^{\t}=\ex[\mathbf{H}(\bY_i; \nnu)], \label{eq:eta.g}\\
    \hat{\boldsymbol{\eta}}(\nnu) &= (\hat\eta_1(\nnu), \dots, \hat\eta_{\ell}(\nnu))^{\t}=\frac{1}{n}\sum_{i=1}^n \mathbf{H}(\bY_i; \nnu). \label{eq:etahat.g}
\end{align}
Substituting the ML estimators for $\nnu$ into Equation \ref{eq:etahat.g}, $\hat\eeta(\hat\nnu)$ serves as an empirical estimator of $\eeta(\nnu)$, subject to sampling variability from both item parameter estimation and the use of sample mean. 

Now, let $\varphi: \mathbb{R}^{\ell} \to \mathbb{R}$ be a differentiable transformation function applied to $\eeta(\nnu)$ and $\hat\eeta(\hat\nnu)$. Later, we express population reliability coefficients as $\varphi{(\eeta(\nnu))}$ with suitable 
 choices of $\varphi$. Similarly, applying $\varphi$ to $\hat\eeta(\hat\nnu)$ results in an estimated reliability coefficient $\varphi{(\hat\eeta(\hat\nnu))}$, which is of primary interest. By first deriving the asymptotic covariance matrix of $\hat\eeta(\hat\nnu)$ and then applying the Delta method \parencite[e.g.,][Lemma 5.3.3]{bickel2015}, we establish the asymptotic normality of $\varphi(\hat{\eeta}(\hat{\nnu}))$ as follows:
\begin{align}
    &\sqrt{n}[ \varphi(\hat{\eeta}(\hat{\nnu})) - \varphi(\eeta(\nnu)) ]     \xrightarrow{d}  \mathcal{N} \left( 0, \nabla\varphi(\eeta(\nnu))^{\t} \bm\Sigma(\nnu) \nabla \varphi(\eeta(\nnu))\right).
    \label{eq:se1}
\end{align}
In Equation \ref{eq:se1}, $\bm{\Sigma}(\nnu)$ denotes the asymptotic covariance matrix of $\hat\eeta(\hat\nnu)$, and $\nabla\varphi$ denotes the \mkr{$\ell \times 1$} Jacobian vector of $\varphi$, assumed to be non-vanishing at $\nnu$. The exact form of $\bm{\Sigma}(\nnu)$ and the details of the derivation are provided in the supplementary document. 
  
The asymptotic SE for $\varphi(\hat{\eeta}(\hat{\nnu}))$ is then obtained by
\begin{equation}
    \mathrm{SE}[\varphi(\hat{\eeta}(\hat{\nnu}))]=\sqrt{\frac{1}{n}\left[\nabla\varphi(\hat{\eeta}(\hat{\nnu}))^{\top}\hat{\bm\Sigma}(\hat{\nnu}) \nabla \varphi(\hat\eeta(\hat\nnu))\right]},
    \label{eq:se_final}
\end{equation}       
in which $\hat{\bm\Sigma}(\hat\nnu)$ is a consistent estimator of ${\bm\Sigma(\nnu)}$.
The approximate $100(1-\alpha)\%$ confidence interval (CI) for the population-level reliability $\varphi(\eeta(\nnu))$ is constructed as 
\begin{equation}
    \varphi(\hat{\eeta}(\hat{\nnu})) \pm z_{1-\alpha/2}\mathrm{SE}[\varphi(\hat{\eeta}(\hat{\nnu}))],
    \label{eq:CI}
\end{equation}       
where $z_{1-\alpha/2}$ denotes the $(1-\alpha/2)$th quantile of the standard normal distribution.

\subsection{PRMSE}

To apply the general theory to PRMSE of $\Theta_i$, define the function $\mathbf{H}$ in Equation \ref{eq:g.H} as 
\begin{equation}
  \mathbf{H}(\bY_i; \nnu)=(H_1(\bY_i; \nnu), H_2(\bY_i; \nnu), H_3(\bY_i; \nnu))^\top,
  \label{eq:H_prmse}
\end{equation}
in which $H_1(\bY_i;\nnu)=\ex(\Theta_i|\bY_i;\nnu)$, $H_2(\bY_i; \nnu)=\ex(\Theta_i|\bY_i;\nnu)^2$, and $H_3(\bY_i; \nnu)=\var(\Theta_i|\bY_i;\nnu)$.
Also, let $\eeta=(\eta_1, \eta_2, \eta_3 )^{\t}$ and $\hat\eeta=(\hat\eta_1, \hat\eta_2, \hat\eta_3 )^{\t}$ be the population expectation and the sample mean of Equation \ref{eq:H_prmse}, respectively, as explained in Equations \ref{eq:eta.g} and \ref{eq:etahat.g}.

Using a transformation function $\varphi_{\rm PRMSE}: \mathbb{R}^{3} \to \mathbb{R}$ such that $\varphi_{\rm PRMSE}(\xx)=(x_2 - x_1^2)/(x_2 - x_1^2 + x_3)$ where $\xx=(x_1, x_2, x_3)^{\t}$, the population PRMSE of $\Theta_i$ (Equation \ref{eq:prmse}) can be re-expressed by 
\begin{align}
   \varphi_{\rm PRMSE}(\eeta(\nnu))
   &=\frac{\eta_2(\nnu) - \eta_1^2(\nnu)}{\eta_2(\nnu) - \eta_1^2(\nnu) + \eta_3(\nnu)}.
   \label{eq:prmse_pop2}
\end{align}
Here, the numerator of Equation \ref{eq:prmse} becomes $\var[\ex(\Theta_i|\bY_i;\nnu);\nnu]=\eta_2(\nnu) - \eta_1^2(\nnu)$ by rewriting the variance in terms of expectations.
The denominator of Equation \ref{eq:prmse} is first decomposed into two terms by the law of 
 total variance, $\var(\Theta_i)=\var[\ex(\Theta_i|\bY_i;\nnu);\nnu]+\ex[\var(\Theta_i|\bY_i;\nnu);\nnu]$, which simplifies using $\eeta(\nnu)$ to $\var(\Theta_i)=\eta_2(\nnu) - \eta_1^2(\nnu) + \eta_3(\nnu)$.
The corresponding sample PRMSE estimate can then be obtained by
\begin{equation}
   \varphi_{\rm PRMSE}(\hat{\eeta}(\hat{\nnu}))
   = \frac{\hat{\eta}_2(\hat\nnu) - \hat{\eta}_1^2(\hat\nnu)}{\hat{\eta}_2(\hat\nnu) - \hat{\eta}_1^2(\hat\nnu) + \hat{\eta}_3(\hat\nnu)}.
   \label{eq:prmse_est}
\end{equation}
The asymptotic SE for $\varphi_{\rm PRMSE}(\hat{\eeta}(\hat{\nnu}))$ and the $100(1-\alpha)\%$ CI covering $\varphi_{\rm PRMSE}(\eeta(\nnu))$ are derived from Equations \ref{eq:se_final} and \ref{eq:CI}, respectively.

\subsection{CTT Reliability}

The SE derivation for CTT reliability is more involved due to the need to approximate an intractable integral. To simplify the notation, let $\tau(\theta_i;\nnu)$ be the true score of the observed score of interest, $s(\bY_i;\nnu)=\ex(\Theta_i|\bY_i;\nnu)$; that is, $\tau(\theta_i;\nnu)=\ex[s(\bY_i;\nnu)|\theta_i;\nnu]$.
Then, the numerator of Equation \ref{eq:cttrel} can be expressed as 
\begin{equation}
    \var[\tau(\Theta_i;\nnu);\nnu]
    = \ex[\tau(\Theta_i;\nnu)^2; \nnu] - \ex[\tau(\Theta_i;\nnu); \nnu]^2.
    \label{eq:ctt_nu}
\end{equation}
The second term on the right-hand side of Equation \ref{eq:ctt_nu} can be reduced to an expectation with respect to $\bY_i$ by the law of iterated expectations:
\begin{equation}    
    \ex[\tau(\Theta_i;\nnu); \nnu]^2=\ex(\ex[s(\bY_i;\nnu)|\Theta_i;\nnu])^2=\ex[s(\bY_i;\nnu);\nnu]^2.
    \label{eq:scd}
\end{equation}
However, the first term on the right-hand side of Equation \ref{eq:ctt_nu} remains an expectation over the continuous LV $\Theta_i$, requiring an approximation of the integral. We proceed to approximate it numerically by quadrature:
\begin{equation}
    \ex[\tau(\Theta_i;\nnu)^2; \nnu] = \int \tau(\theta_i;\nnu)^2 \phi(\theta_i)d\theta_i \approx\sum_{q=1}^Q \tau(\theta_{iq};\nnu)^2 w_q,
    \label{eq:approx}
\end{equation}
in which $w_q=[\sum_{q=1}^Q \phi(\theta_{iq})]^{-1}\phi(\theta_{iq})$, $q=1,\dots, Q$, are normalized rectangular quadrature weights.
To make Equation \ref{eq:approx} estimable using sample data, we further re-express the conditional expectation $\tau(\theta_i;\nnu)$ as an expectation with respect to the marginal distribution of $\bY_i$:
\begin{equation}  
    \tau(\theta_i;\nnu)=\sum_{\yy_i}s(\yy_i;\nnu)f(\yy_i|\theta_i;\nnu)=\ex\left[ s(\bY_i;\nnu)\frac{f(\bY_i|\theta_i;\nnu)}{f(\bY_i;\nnu)}\right],
    \label{eq:t}
\end{equation}
in which the last equality follows from Bayes' rule. Substituting Equation \ref{eq:t} into the right-hand side of Equation \ref{eq:approx} gives
\begin{equation}
    \ex[\tau(\Theta_i;\nnu)^2; \nnu] \approx \sum_{q=1}^Q \left(\ex\left[ s(\bY_i;\nnu)\frac{f(\bY_i|\theta_{iq};\nnu)}{f(\bY_i;\nnu)}\right]\right)^2 w_q.
    \label{eq:approx2}
\end{equation}

Now, to apply the general theory to CTT reliability, we identify $\mathbf{H}$ in Equation \ref{eq:g.H} as
\begin{equation}
    \mathbf{H}(\bY_i; \nnu)=(H_1(\bY_i; \nnu), H_2(\bY_i; \nnu), H_3(\bY_i; \nnu), \dots, H_{2+Q}(\bY_i; \nnu))^\top,
    \label{eq:H_ctt}
\end{equation}
in which $H_1(\bY_i;\nnu)=s(\bY_i; \nnu)$ and $H_2(\bY_i; \nnu)=s(\bY_i; \nnu)^2$ as in PRMSE. The remaining functions pertain to the use of quadratures and are defined as 
\begin{equation}
    H_{2+q}(\bY_i; \nnu)=H_1(\bY_i; \nnu)\frac{f(\bY_i|\theta_{iq}; \nnu)}{f(\bY_i; \nnu)},\,\,  
    q=1, \dots, Q.
    \label{eq:Hq}
\end{equation}
Let $\eeta=(\eta_1, \dots, \eta_{2+Q} )^{\t}$ and $\hat\eeta=(\hat\eta_1, \dots, \hat\eta_{2+Q} )^{\t}$ be the population expectation and the sample mean of Equation \ref{eq:H_ctt}, respectively, as shown in Equations \ref{eq:eta.g} and \ref{eq:etahat.g}.

Then, by applying a transformation function $\varphi_{\rm Rel}: \mathbb{R}^{2+Q} \to \mathbb{R}$ such that $\varphi_{\rm Rel}(\xx)=(\sum_{q=1}^Q x_{2+q}^2w_q - x_1^2)/(x_2 - x_1^2)$ where $\xx=(x_1, \dots, x_{2+Q})^{\t}$, the population CTT reliability in Equation \ref{eq:cttrel} can be re-expressed by 
\begin{equation}
    \varphi_{\rm Rel}(\eeta(\nnu))
    = \frac{\sum_{q=1}^Q \eta^2_{2+q}(\nnu) w_q - \eta_1^2(\nnu)}{\eta_2(\nnu) - \eta_1^2(\nnu)}.
    \label{eq:ctt_pop2}
\end{equation}
Here, the numerator follows the form presented in Equation \ref{eq:ctt_nu}. The denominator is simply the variance of the EAP score, $\var[s(\bY_i;\nnu);\nnu]$, which also appears as the numerator in Equation \ref{eq:prmse_pop2}. 
The corresponding CTT reliability estimate can then be obtained by
\begin{equation}
    \varphi_{\rm Rel}(\hat{\eeta}(\hat{\nnu}))
    = \frac{\sum_{q=1}^Q \hat\eta^2_{2+q}(\hat\nnu) w_q - \hat{\eta}_1^2(\hat\nnu)}{\hat{\eta}_2(\hat\nnu) - \hat{\eta}_1^2(\hat\nnu)},
    \label{eq:ctt_est}
\end{equation}
with its asymptotic SE given by Equation \ref{eq:se_final}. The $100(1-\alpha)\%$ CI for $\varphi_{\rm Rel}(\eeta(\nnu))$ is constructed by Equation \ref{eq:CI}.

\section{Simulation Study}

\subsection{Simulation Setup}

A simulation study was conducted to examine the finite sample properties of the derived asymptotic SEs and CIs for the two reliability coefficients under the 2PL model (i.e., $K=2$) and the GRM with five response categories (i.e., $K=5$). For the 2PL model, item slope parameters were randomly sampled from $a_j \sim \mathcal{U}[0.5, 2]$ and difficulty parameters from $b_{j1} \sim \mathcal{N}(0, 1)$, following \textcite{andersson2018}. The range specified for the uniform distribution covers approximately 90\% of the slopes estimated from the empirical data presented later in the manuscript. For the GRM, the same distribution was used for the slope parameters, while the intercept parameters were generated following \textcite{liu2014comparing}. Specifically, the difficulty parameter for the first category was randomly sampled from $b_{j1} \sim \mathcal{N}(-1.5, 0.5^2)$, and higher difficulties were obtained as $b_{jk} = b_{j,k-1} + d$ for $k = 2, 3, 4$, where the increment was drawn from $d \sim \mathcal{N}(1, 0.2^2)$. The intercepts were then computed as $c_{jk}= -a_jb_{jk}$. The LV distribution was assumed to be $\mathcal{N}(0, 1)$ for both models.

For each model and each reliability coefficient, nine simulation conditions were determined by fully crossing two factors: (1) Sample size ($n=250, 500, 1,000$) and (2) test length ($m=8, 16, 32$ for the 2PL model and $m=4, 8, 16$ for the GRM). A sample size of 500 or larger is generally recommended to obtain stable item parameter estimates for the 2PL model \parencite[e.g.,][Chapter 5]{de2013theory}. Using $n=500$ as a reference for a moderate sample size condition, we selected $n=250$ to represent a small-sample scenario in which estimation may be less stable. Additionally, we included $n=1,000$ as a large-sample condition to assess whether the reliability estimators exhibit the asymptotic properties derived in this study. 

Test lengths were determined to match reliability levels of substantive interest. We first selected target reliability levels of 0.70, 0.80, and 0.90 to cover a range from minimally acceptable to excellent measurement precision. The value of 0.70 is commonly regarded as the lower bound of acceptability, 0.80 as a benchmark for research applications, and 0.90 or higher as indicative of excellent measurement precision \parencite[e.g.,][]{cicchetti1994guidelines}. Test lengths for each model were then chosen to approximate these target values under the data-generating conditions, which also align with the number of items commonly observed in psychological and educational assessments. We note that the current simulation design unavoidably confounds test length with reliability level because the slope parameters are restricted to a fixed, realistic range. To respond to a referee's request and to make the study self-contained, we report additional simulation results in the supplementary document, in which we attempted to disentangle the effects of test length and reliability. These additional simulations used different ranges for the slope parameters.

Under each simulation condition, 500 datasets were generated. Table \ref{tb:pop_rel} presents the population-level PRMSE and CTT reliability under the 2PL model and the GRM for each test length, computed using the true item parameters based on Equations \ref{eq:prmse_pop2} and \ref{eq:ctt_pop2}, respectively.

\begin{table}[h]
\caption{Population-level PRMSE and CTT reliability under the 2PL model and GRM for each test length. The values were obtained using true item parameters and population moment calculations. 
\\Note. For conditions marked with asterisk, the values were approximated using one million Monte Carlo sample due to the large number of possible response patterns ($2^{32}=4,294,967,296$ for the 2PL model and $5^{16}=152,587,890,625$ for the GRM).}
    \label{tb:pop_rel}
\centering
\begin{tabular}{c c c c}
\toprule
Model & Test Length & PRMSE & CTT Reliability \\
\midrule

\multirow{3}{*}{2PL} 
  & 8  & 0.700 & 0.717 \\
  & 16 & 0.801 & 0.808 \\
  & 32$^{\ast}$ & 0.891 & 0.896  \\

\midrule

\multirow{3}{*}{GRM} 
  & 4  & 0.697 & 0.699 \\
  & 8  & 0.801 & 0.804 \\
  & 16$^{\ast}$ & 0.889 & 0.891  \\

\bottomrule
\end{tabular}
\end{table}

Upon data generation, the \texttt{mirt} package version 1.44.0 \parencite{chalmers2012mirt} in R \parencite{R} was used for parameter estimation. We adopted \texttt{mirt}'s default setting for the expectation-maximization algorithm \parencite{bock1981marginal}. More specifically, 61 equally spaced rectangular quadrature points, ranging from $-6$ to 6, were used to approximate the marginal likelihood function in Equation \ref{eq:marginal}, which matches the number of quadrature points we used to approximate the integral in Equation \ref{eq:approx} (i.e., $Q=61$). To obtain the information matrix in Equation \ref{eq:info}, we calculated the observed information matrix by setting \texttt{SE.type=`Louis'} in \texttt{mirt}.

After fitting the model, the PRMSE and CTT reliability estimates were computed along with their asymptotic SEs. For both types of reliability coefficients, the means and standard deviations (SDs) of the point estimates across 500 replications were recorded. The average of the SEs estimated using our formula was then compared with the empirical SD, serving as a criterion for evaluating the accuracy of the SE estimation. The empirical coverage rate of the 95$\%$ CI was also examined.

With the default convergence criteria, non-convergence occurred in one out of the 500 datasets under the condition with the smallest sample size ($n=250$) and shortest test length ($m=8$) for the 2PL model. This non-convergent case was excluded from the analysis. Under the condition with the smallest sample size ($n = 250$) and the longest test length ($m = 16$) for the GRM, the estimated CTT reliability coefficient exceeded one in 14 datasets. This is because the numerator and denominator of the CTT reliability coefficient are calculated separately from the sample (Equation \ref{eq:ctt_est}); namely, the sample-based reliability estimate may exceed its theoretical upper bound of one due to sampling variability. Fortunately, this issue arose only when the sample size was $n=250$ under high reliability condition, with an occurrence rate less than 3\%. Meanwhile, our estimator of PRMSE never exceeds one because, as shown in Equation \ref{eq:prmse_est}, the denominator is always greater than or equal to the numerator. For CTT reliability, constructing a normalized estimator requires a more complex transformation function (i.e., $\varphi_{\rm Rel}$). For completeness, we provide derivations and additional simulation results about this alternative estimator of CTT reliability in the supplementary document.   

\subsection{Results: PRMSE}

Tables \ref{tb:prmse_2pl_final} and \ref{tb:prmse_grm_final} present the simulation results for PRMSE under the 2PL model and the GRM, respectively. As the results under the two models closely mirror each other, we provide a detailed discussion on the results with the 2PL model and omit further explanation for the GRM.

First, before evaluating the accuracy of the SE estimates, we examined the accuracy of the point estimates by comparing the third and forth columns of Table \ref{tb:prmse_2pl_final}. The results indicate a slight overestimation of PRMSE when the sample size is small, with this bias diminishing as the sample size increases. Specifically, the relative bias for the $n=250$ condition---computed as the difference between the estimated and true values divided by the true values---is 0.011, 0.006, and 0.002 for test lengths of 8, 16, and 32, respectively. Within each test length, these values further decrease for $n=500$ and $n=1,000$, in alignment with asymptotic theory.

Next, the average of the asymptotic SE estimates (the sixth column of Table \ref{tb:prmse_2pl_final}) was compared with the empirical SDs of the PRMSE estimates (the fifth column of Table \ref{tb:prmse_2pl_final}). Across all conditions, the differences between the two were negligible, with a discrepancy no more than 0.002. These results verify that the derived asymptotic SE formula provides a close approximation to the true sampling variability of PRMSE, even for a sample size as small as $n=250$. 

Finally, the empirical coverage rates of the 95$\%$ CIs, presented in the seventh column of Table \ref{tb:prmse_2pl_final}, were evaluated. Coverage rates outside the Monte Carlo error bounds\footnote{The Monte Carlo error bounds were computed from a normal approximation to the binomial distribution: $0.95 \pm 1.96\sqrt{0.95(1-0.95)/\text{500}} \approx [0.931, 0.969]$.} are shown in bold in the table. The last two columns report the mean lower and upper bounds of the CIs. A notable pattern found is the undercoverage observed for $n=250$ across all test length conditions. Given that the SE estimates tend to be accurate, this undercoverage is likely due to poor normal approximation and bias in the point estimates under small samples. Within each test length, the coverage rates improve as the sample size increases, consistent with the improved accuracy of the point estimates under larger samples. However, the results for the $n=250$ condition across different test lengths show that improvements in the point estimates (with increasing test lengths) do not necessarily result in improved coverage. This is because the sampling distributions remain noticeably skewed under the $n=250$ condition, regardless of the magnitude of the point estimate bias. This skewness diminishes as the sample size increases, leading to better recovery of the target coverage.

\begin{table}[t]
\centering
\begin{tabular}{ccccccccc}
\toprule
$n$ & $m$ & PRMSE True & PRMSE Est. & Emp. SD & SE & Coverage & LB & UB \\
\midrule
\multirow{3}{*}{250} 
 & 8  & 0.700 & 0.708 & 0.023 & 0.024 & \textbf{0.920} & 0.662 & 0.754 \\
 & 16 & 0.801 & 0.806 & 0.015 & 0.015 & \textbf{0.916} & 0.777 & 0.836 \\
 & 32 & 0.891 & 0.893 & 0.007 & 0.007 & \textbf{0.920} & 0.879 & 0.907 \\
\midrule
\multirow{3}{*}{500} 
 & 8  & 0.700 & 0.703 & 0.017 & 0.017 & 0.940 & 0.669 & 0.736 \\
 & 16 & 0.801 & 0.804 & 0.010 & 0.011 & 0.936 & 0.783 & 0.825 \\
 & 32 & 0.891 & 0.892 & 0.005 & 0.005 & 0.946 & 0.883 & 0.902 \\
\midrule
\multirow{3}{*}{1,000} 
 & 8  & 0.700 & 0.701 & 0.012 & 0.012 & 0.950 & 0.678 & 0.725 \\
 & 16 & 0.801 & 0.802 & 0.008 & 0.008 & 0.948 & 0.787 & 0.817 \\
 & 32 & 0.891 & 0.892 & 0.003 & 0.004 & 0.960 & 0.885 & 0.898 \\
\bottomrule
\end{tabular}

\caption{Simulation results for PRMSE under the 2PL model. $n$: Sample size. $m$: Test length. PRMSE True: Population-level PRMSE obtained using true item parameters and population moment calculations. PRMSE Est.: Mean of the PRMSE estimates across replications. Emp. SD: Empirical standard deviation of the point estimates. SE: Mean of the asymptotic SEs estimated by the derived formula. Coverage: Empirical coverage rate of the 95$\%$ CI; values falling outside the Monte Carlo error bounds are displayed in bold. LB: Mean lower bound of the CI. UB: Mean upper bound of the CI.}
\label{tb:prmse_2pl_final}
\end{table}

\begin{table}[h]
\centering
\begin{tabular}{ccccccccc}
\toprule
$n$ & $m$ & PRMSE True & PRMSE Est. & Emp. SD & SE & Coverage & LB & UB \\
\midrule
\multirow{3}{*}{250} 
 & 4  & 0.697 & 0.707 & 0.037 & 0.035 & 0.934 & 0.638 & 0.776 \\
 & 8  & 0.801 & 0.804 & 0.018 & 0.018 & \textbf{0.922} & 0.770 & 0.839 \\
 & 16 & 0.889 & 0.891 & 0.010 & 0.009 & \textbf{0.918} & 0.872 & 0.909 \\
\midrule
\multirow{3}{*}{500} 
 & 4  & 0.697 & 0.700 & 0.025 & 0.024 & 0.954 & 0.652 & 0.748 \\
 & 8  & 0.801 & 0.804 & 0.012 & 0.012 & 0.932 & 0.779 & 0.828 \\
 & 16 & 0.889 & 0.889 & 0.007 & 0.007 & 0.942 & 0.876 & 0.903 \\
\midrule
\multirow{3}{*}{1,000} 
 & 4  & 0.697 & 0.699 & 0.017 & 0.017 & 0.942 & 0.666 & 0.732 \\
 & 8  & 0.801 & 0.801 & 0.009 & 0.009 & 0.954 & 0.784 & 0.819 \\
 & 16 & 0.889 & 0.890 & 0.005 & 0.005 & 0.950 & 0.880 & 0.899 \\
\bottomrule
\end{tabular}

\caption{Simulation results for PRMSE under the GRM. $n$: Sample size. $m$: Test length. PRMSE True: Population-level PRMSE obtained using true item parameters and population moment calculations. PRMSE Est.: Mean of the PRMSE estimates across replications. Emp. SD: Empirical standard deviation of the point estimates. SE: Mean of the asymptotic SEs estimated by the derived formula. Coverage: Empirical coverage rate of the 95\% CI; values falling outside the Monte Carlo error bounds are displayed in bold. LB: Mean lower bound of the CI. UB: Mean upper bound of the CI.}
\label{tb:prmse_grm_final}
\end{table}

\subsection{Results: CTT Reliability}

Tables \ref{tb:cttrel_2pl_final} and \ref{tb:cttrel_grm_final} present the simulation results for CTT reliability under the 2PL model and the GRM, respectively. Because the results across two models were similar, we provide a detailed discussion of the findings for the 2PL model and highlight only a few distinct observations for the GRM. 

First, the accuracy of the reliability point estimates was evaluated by comparing the third and forth columns of Table \ref{tb:cttrel_2pl_final}. The results indicate that CTT reliability tends to be overestimated, particularly when the sample size is small and the test length is long. When we consider the fixed test length, the degree of the overestimation diminishes as the sample size increases, consistent with the asymptotic theory. Specifically, the relative bias for the $n=250$ condition---computed as the difference between the estimated and true values divided by the true values---is 0.018, 0.016, and 0.023 for test lengths of 8, 16, and 32, respectively. These values decrease to 0.003, 0.004, and 0.008, respectively, in the $n=1,000$ condition. Notably, under the longest test length condition ($m=32$), we observed several upward outliers that produce a positively skewed sampling distribution. The larger relative bias in the condition could be attributable to these outliers. 

\begin{table}[t]
\centering
\begin{tabular}{ccccccccc}
\toprule
$n$ & $m$ & Rel. True & Rel. Est. & Emp. SD & SE & Coverage & LB & UB \\
\midrule
\multirow{3}{*}{250} 
 & 8  & 0.717 & 0.730 & 0.026 & 0.027 & \textbf{0.908} & 0.677 & 0.784 \\
 & 16 & 0.808 & 0.821 & 0.018 & 0.019 & \textbf{0.920} & 0.784 & 0.859 \\
 & 32 & 0.896 & 0.915 & 0.016 & 0.020 & 0.952 & 0.876 & 0.953 \\
\midrule
\multirow{3}{*}{500} 
 & 8  & 0.717 & 0.722 & 0.018 & 0.019 & 0.938 & 0.684 & 0.759 \\
 & 16 & 0.808 & 0.815 & 0.012 & 0.013 & 0.944 & 0.789 & 0.841 \\
 & 32 & 0.896 & 0.905 & 0.009 & 0.011 & 0.966 & 0.884 & 0.926 \\
\midrule
\multirow{3}{*}{1,000} 
 & 8  & 0.717 & 0.719 & 0.013 & 0.014 & 0.948 & 0.693 & 0.746 \\
 & 16 & 0.808 & 0.811 & 0.009 & 0.009 & 0.946 & 0.793 & 0.829 \\
 & 32 & 0.896 & 0.901 & 0.006 & 0.007 & 0.964 & 0.887 & 0.914 \\
\bottomrule
\end{tabular}

\caption{Simulation results for CTT reliability under the 2PL model. $n$: Sample size. $m$: Test length. Rel. True: Population-level CTT reliability obtained using true item parameters and population moment calculations. Rel. Est.: Mean of the CTT reliability estimates across replications. Emp. SD: Empirical standard deviation of the point estimates. SE: Mean of the asymptotic SEs estimated by the derived formula. Coverage: Empirical coverage rate of the 95$\%$ CI; values falling outside the Monte Carlo error bounds are displayed in bold. LB: Mean lower bound of the CI. UB: Mean upper bound of the CI.}
\label{tb:cttrel_2pl_final}
\end{table}

\begin{table}[h]
\centering
\begin{tabular}{ccccccccc}
\toprule
$n$ & $m$ & Rel. True & Rel. Est. & Emp. SD & SE & Coverage & LB & UB \\
\midrule
\multirow{3}{*}{250} 
 & 4  & 0.699 & 0.717 & 0.038 & 0.038 & \textbf{0.910} & 0.643 & 0.791 \\
 & 8  & 0.804 & 0.820 & 0.022 & 0.024 & \textbf{0.924} & 0.772 & 0.867 \\
 & 16 & 0.891 & 0.929 & 0.030 & 0.038 & 0.938 & 0.854 & 1.004 \\
\midrule
\multirow{3}{*}{500} 
 & 4  & 0.699 & 0.706 & 0.025 & 0.026 & 0.946 & 0.656 & 0.757 \\
 & 8  & 0.804 & 0.813 & 0.014 & 0.015 & \textbf{0.926} & 0.783 & 0.844 \\
 & 16 & 0.891 & 0.909 & 0.015 & 0.019 & 0.946 & 0.871 & 0.947 \\
\midrule
\multirow{3}{*}{1,000} 
 & 4  & 0.699 & 0.703 & 0.017 & 0.018 & 0.936 & 0.669 & 0.738 \\
 & 8  & 0.804 & 0.808 & 0.010 & 0.010 & 0.952 & 0.787 & 0.828 \\
 & 16 & 0.891 & 0.900 & 0.008 & 0.010 & 0.962 & 0.880 & 0.919 \\
\bottomrule
\end{tabular}

\caption{Simulation results for CTT reliability under the GRM. $n$: Sample size. $m$: Test length. Rel. True: Population-level CTT reliability obtained using true item parameters and population moment calculations. Rel. Est.: Mean of the CTT reliability estimates across replications. Emp. SD: Empirical standard deviation of the point estimates. SE: Mean of the asymptotic SEs estimated by the derived formula. Coverage: Empirical coverage rate of the 95$\%$ CI; values falling outside the Monte Carlo error bounds are displayed in bold. LB: Mean lower bound of the CI. UB: Mean upper bound of the CI.}
\label{tb:cttrel_grm_final}
\end{table}

Next, the average asymptotic SE (the sixth column of Table \ref{tb:cttrel_2pl_final}) was compared with the empirical SD of the CTT reliability estimates (the fifth column of Table \ref{tb:cttrel_2pl_final}). Except for the condition with $m=32$, the differences between the two were negligible, with discrepancies no larger than 0.002. Only under the $m=32$ condition with the smallest sample size ($n=250$) was the SE noticeably overestimated compared to the empirical SD, showing a discrepancy of 0.004. However, consistent with asymptotic theory, this discrepancy diminishes to a negligible level as the sample size increases.

Finally, the empirical coverage rates of the 95$\%$ CIs, shown in the seventh column of Table \ref{tb:cttrel_2pl_final}, were evaluated, with values outside the Monte Carlo error bounds highlighted in bold. Similar to the PRMSE results, the coverage tended to exhibit undercoverage in the small-sample condition ($n=250$). However, for the $m=32$ condition where both the point estimate and SE were the least accurate, the coverage remained close to the nominal level. This pattern indicates that coverage is determined by the interaction of multiple factors. In this case, the combination of a positively skewed yet still relatively concentrated sampling distribution and the wider CIs produced by overestimated SEs may have jointly contributed to the adequate coverage. Importantly, both the normal approximation of the sampling distribution and the accuracy of the SE estimates improve as the sample size increases to $n=500$ and $n=1,000$, yielding stable recovery of the nominal coverage level in larger samples across all test lengths.

For the GRM, the overall pattern of results was similar to those in the 2PL model. One slight distinction was that under the longest test-length condition ($m=16$) with the smallest sample size ($n=250$), the positive skewness of the sampling distribution was more pronounced than in the corresponding condition under the 2PL model. This phenomenon is reflected in the third row of Table \ref{tb:cttrel_grm_final} as the substantially overestimated point estimate, along with the large empirical SD and average SE values. Fourteen reliability estimates exceeded one in this condition (as explained prior to the result section), which may have contributed to these exacerbated findings. The positive skewness remained salient at $n=500$ and persisted, though to a lesser degree, at $n=1000$. Nevertheless, the coverage stays close to the nominal level across all sample sizes, consistent with the observation under the 2PL model.

\section{Empirical Example}

In this section, we present an empirical illustration for reporting reliability and PRMSE and quantifying their sampling variability. The analyses were based on the ``SAT12'' data set available from the \texttt{mirt} package \parencite{chalmers2012mirt}, which contain responses from 600 students to 32 dichotomous items from a grade 12 science assessment test covering chemistry, biology, and physics \parencite[][p. 200]{chalmers2012mirt}. R code for the analyses is provided in the supplementary material.

\subsection{CTT Reliability vs. PRMSE}

Consider a scenario where a researcher estimates EAP scores to measure latent science proficiency (i.e. the LV) and wishes to evaluate how well these observed EAP scores reflect the underling latent science proficiency. In this case, the appropriate reliability coefficient to report is the CTT reliability of the EAP score.\footnote{It should be noted that CTT reliability is a property of the observed score (Equation \ref{eq:cttrel}), and different types of observed score yield different reliability coefficients.} 
Using the SAT12 data, CTT reliability for the EAP score, calculated based on Equation \ref{eq:ctt_est}, was found to be 0.918, indicating that 91.8\% of the variance in the observed EAP score is explained by individual differences in the latent proficiency. This is equivalent to state that 8.2\% of the variance in the observed EAP score is attributed to measurement error. 

Conversely, suppose that the research interest begins with the unobservable science proficiency itself, and the goal is to evaluate how well this latent proficiency can be predicted from the 32 items.
The appropriate reliability coefficient to report in this case is the PRMSE for the LV. 
For the SAT12 data, the PRMSE for the LV, computed based on Equation \ref{eq:prmse_est}, was found to be 0.838, indicating that 83.8\% of the variance in the latent science proficiency is accounted for by the responses to the 32 items.\footnote{This value can also be obtained using the \texttt{mirt} package by calling the ``empirical reliability'', as demonstrated in the provided R code.} The remaining 16.2\% quantifies prediction error in this case.

\subsection{Quantification of Sampling Variability}

With a test length of 32, computing population-level moments over all possible response patterns is computationally intractable, and therefore, the values of the CTT reliability and PRMSE coefficients (0.918 and 0.838, respectively) were estimated using sample moments. Because these values contain sampling variability from item parameter estimation and the use of sample moments, the uncertainty should be quantified and reported along with the reliability point estimates, just as it is standard practice to report item parameter estimates with their standard errors. Applying the formulas derived in our study, the SE of the CTT reliability was found to be 0.036, yielding a 95\% CI of [0.847, 0.990]. For the PRMSE, the SE was found to be 0.009, producing a 95\% CI of [0.821, 0.856].

\section{Discussion}

As a key index of measurement precision, reliability coefficients are reported in nearly all psychological and educational research involving latent constructs. However, these coefficients are inherently subject to sampling variability. Existing approaches to quantifying this variability have been limited to cases where item parameter estimation is the only source of uncertainty. Unlike previous studies, we focus on situations where reliability coefficients are computed using sample moments in place of population moments, a scenario that is typically unavoidable when estimating CTT reliability and PRMSE for long tests. Our work contributes to the literature in four ways. First, we introduce a general framework for deriving SEs that account for the two sources of variability simultaneously. Second, we provide full SE derivations for two specific examples: CTT reliability for the EAP score and PRSME for the LV. Third, although not emphasized in the main text, our approach also enables the estimation of the exact forms of CTT reliability in long tests for any observed score by re-expressing the true score (i.e., the conditional expectation) as a marginal expectation, as shown in Equation \ref{eq:t}. The marginal expectation can then be estimated by using sample means. Approximating the true score by sample moments eliminates the need to compute expectations over all possible response patterns and enables direct application of our general SE formula. Finally, our SE formula can potentially be used for sample-size planning in reliability studies.

The key findings of our simulation study are summarized as follows. First, with a small sample size, point estimates for both CTT reliability of the EAP score and PRMSE of the LV tend to be inaccurate with skewed sampling distribution, leading to suboptimal CI coverage. However, as the sample size increases, both point estimates and SEs closely align with the target values, and the coverage rates also reach the nominal level. These results suggest that the derived SE formulas precisely characterize the sampling variability and therefore can serve as a valid uncertainty quantification measure in moderate to large samples under commonly used test lengths.

There are several avenues for future research that extend beyond the scope of the current work. First, our derivations focused on cases where the EAP score of the LV is used as the observed score in CTT reliability, and the untransformed LV is used as the latent score in PRMSE. Future work could extend our derivations to accommodate CTT reliability for other types of observed scores and PRMSE for other types of latent scores. Different scores require different formulations of the function $H$ (Equations \ref{eq:H_prmse} and \ref{eq:H_ctt}) and its gradient with respect to model parameters (see Section A of the supplementary document for details). Beyond CTT reliability and PRMSE, the SE calculation is also needed for other indices of measurement precision that are more broadly defined by the association between latent and observed scores \parencite{liu2025general}, offering another potential direction for extension.

Second, while our study focused on the unidimensional GRM and assumed a standard normal distribution for the LV, future work could extend the derivations to more complex measurement models. The general theory we propose remains applicable as long as a fully specified parametric model is assumed. One promising direction is the extension to multidimensional IRT models, under which the SE derivations for reliability coefficients have not yet been explored in the literature. The extension for the case of CTT reliability may pose challenges due to the intractable integrations involved in computing the mean true score; meanwhile, the extension for PRMSE should be straightforward.

Third, throughout the article, we assumed that the model is correctly specified. This assumption is mild since reliability calculation is model-based in nature. Goodness-of-fit assessment for IRT models is important and has been extensively studied in the literature \parencite[e.g.,][]{maydeu2005limited,maydeu2006limited,joe2010general}. Specifically for the normality of the LV, formal tests have been developed in, for example, \textcite{monroe2021testing} and \textcite{Sung_2025}. When the model is found to be misspecified, it is generally not recommended to proceed with any model-based inference. However, from a practical standpoint, it would still be valuable to examine the performance of our method under conditions of close fit \parencite{maydeu2014assessing}. 

Fourth, our simulation study revealed consistent overestimation in point estimates for both CTT reliability and PRMSE when the sample size was small. This finding is consistent with \textcite{andersson2018}, who reported similar results for marginal and test reliability coefficients. Future research could explore bias-correction methods to improve the performance of the current approach. For instance, \textcite{andersson2018} suggested a nonparametric bootstrap method \parencite{davison1997bootstrap} to estimate the bias, which can then be used to construct bias-adjusted CIs. Another potential direction is to apply suitable transformation (e.g., Fisher $z$-transformation), which may improve the normal approximation and therefore help achieve more accurate coverage in small samples. This transformation may also help address the poor performance of Wald-type CIs when the true parameters are near the boundary of the parameter space.

Lastly, our work analytically derived the asymptotic SEs for IRT reliability coefficients, but alternative approaches, such as the one based on simulation suggested in \textcite{liu2024}, could be explored. Developing methods that do not rely on the large-sample based normal approximation could be an another direction for future research.

\printbibliography



\section*{Supplementary material}

# A Derivations of Asymptotic Standard Errors

## A.1 General Theory

To derive the asymptotic standard error (SE) of $\varphi(\hat{\boldsymbol{\eta}}(\hat{\boldsymbol{\nu}}))$, we first study the asymptotic behavior of $\hat{\boldsymbol{\eta}}(\hat{\boldsymbol{\nu}})$ before any transformation. Rewrite

$$\sqrt{n}\left[\hat{\boldsymbol{\eta}}(\hat{\boldsymbol{\nu}}) - \boldsymbol{\eta}(\boldsymbol{\nu})\right]$$
$$= \sqrt{n}\left[\hat{\boldsymbol{\eta}}(\hat{\boldsymbol{\nu}}) - \hat{\boldsymbol{\eta}}(\boldsymbol{\nu})\right] + \sqrt{n}\left[\hat{\boldsymbol{\eta}}(\boldsymbol{\nu}) - \boldsymbol{\eta}(\boldsymbol{\nu})\right]. \tag{S1}$$

Let $p_n(\mathbf{y}_i)$ and $f(\mathbf{y}_i; \boldsymbol{\nu})$ be the sample proportion and model-implied probability of response pattern $\mathbf{y}_i$. Using these notations, the first term on the right-hand side of Equation S1 is expressed as

$$\sqrt{n}\left[\hat{\boldsymbol{\eta}}(\hat{\boldsymbol{\nu}}) - \hat{\boldsymbol{\eta}}(\boldsymbol{\nu})\right]$$
$$= \sqrt{n}\sum_{\mathbf{y}_i} p_n(\mathbf{y}_i)\left[\mathbf{H}(\mathbf{y}_i; \hat{\boldsymbol{\nu}}) - \mathbf{H}(\mathbf{y}_i; \boldsymbol{\nu})\right]$$
$$= \underbrace{\sqrt{n}\sum_{\mathbf{y}_i}\left[p_n(\mathbf{y}_i) - f(\mathbf{y}_i; \boldsymbol{\nu})\right]\left[\mathbf{H}(\mathbf{y}_i; \hat{\boldsymbol{\nu}}) - \mathbf{H}(\mathbf{y}_i; \boldsymbol{\nu})\right]}_{(I)}$$
$$+ \underbrace{\sqrt{n}\sum_{\mathbf{y}_i} f(\mathbf{y}_i; \boldsymbol{\nu})\left[\mathbf{H}(\mathbf{y}_i; \hat{\boldsymbol{\nu}}) - \mathbf{H}(\mathbf{y}_i; \boldsymbol{\nu})\right]}_{(II)}. \tag{S2}$$

(I) in Equation S2 is $o_p(1)$ because (a) for each $\mathbf{y}_i$, $\sqrt{n}[p_n(\mathbf{y}_i) - f(\mathbf{y}_i; \boldsymbol{\nu})] = O_p(1)$ as $n \to \infty$ by the asymptotic normality of sample proportions, (b) $\hat{\boldsymbol{\nu}} \xrightarrow{p} \boldsymbol{\nu}$ as $n \to \infty$ by the consistency of the maximum likelihood (ML) estimator, and (c) $\mathbf{H}$ is continuous in $\boldsymbol{\nu}$. Now applying the Delta method to (II) gives

$$(II) = \sum_{\mathbf{y}_i} f(\mathbf{y}_i; \boldsymbol{\nu})\nabla_{\boldsymbol{\nu}}\mathbf{H}(\mathbf{y}_i; \boldsymbol{\nu})\sqrt{n}(\hat{\boldsymbol{\nu}} - \boldsymbol{\nu})$$
$$= \mathbb{E}\left[\nabla_{\boldsymbol{\nu}}\mathbf{H}(\mathbf{Y}_i; \boldsymbol{\nu})\right]^{\top} \mathcal{I}^{-1}(\boldsymbol{\nu})\sqrt{n}\nabla_{\boldsymbol{\nu}}\hat{\ell}(\boldsymbol{\nu}) + o_p(1), \tag{S3}$$

in which $\nabla_{\boldsymbol{\nu}}\mathbf{H}$ produces a $2m \times k$ gradient matrix for $\mathbf{H} \in \mathbb{R}^k$ and $\nabla_{\boldsymbol{\nu}}\hat{\ell}(\boldsymbol{\nu}) = n^{-1}\sum_{i=1}^{n}\nabla_{\boldsymbol{\nu}}\log f(\mathbf{Y}_i; \boldsymbol{\nu})$ from Equation 4 of the main document. Combining

Equations S2 and S3 yields a further expression of Equation S1:

$$\sqrt{n}\left[\hat{\boldsymbol{\eta}}(\hat{\boldsymbol{\nu}}) - \boldsymbol{\eta}(\boldsymbol{\nu})\right]$$
$$= \left(\mathbb{E}[\nabla_{\boldsymbol{\nu}}\mathbf{H}(\mathbf{Y}_i;\boldsymbol{\nu})]^{\top}\mathcal{I}^{-1}(\boldsymbol{\nu}) : \mathbf{I}_{k\times k}\right)\sqrt{n}\begin{pmatrix}\nabla_{\boldsymbol{\nu}}\hat{\ell}(\boldsymbol{\nu}) \\ \hat{\boldsymbol{\eta}}(\boldsymbol{\nu}) - \boldsymbol{\eta}(\boldsymbol{\nu})\end{pmatrix} + o_p(1)$$
$$\xrightarrow{d} \mathcal{N}\left(\mathbf{0}, \boldsymbol{\Sigma}(\boldsymbol{\nu})\right), \tag{S4}$$

in which
$$\boldsymbol{\Sigma}(\boldsymbol{\nu}) = \mathbb{E}[\nabla_{\boldsymbol{\nu}}\mathbf{H}(\mathbf{Y}_i;\boldsymbol{\nu})]^{\top}\mathcal{I}^{-1}(\boldsymbol{\nu})\boldsymbol{\Omega}(\boldsymbol{\nu})\mathcal{I}^{-1}(\boldsymbol{\nu})\mathbb{E}[\nabla_{\boldsymbol{\nu}}\mathbf{H}(\mathbf{Y}_i;\boldsymbol{\nu})]. \tag{S5}$$

Note that the colon (:) in the second line of Equation S4 denotes column-wise concatenation of matrix blocks. Additionally, $\boldsymbol{\Omega}(\boldsymbol{\nu})$ in Equation S5 denotes the covariance matrix of $\nabla_{\boldsymbol{\nu}}\hat{\ell}(\boldsymbol{\nu})$ and $\hat{\boldsymbol{\eta}}(\boldsymbol{\nu})$:

$$\boldsymbol{\Omega}(\boldsymbol{\nu}) = \begin{pmatrix}\mathcal{I}(\boldsymbol{\nu}) & \mathbf{A}(\boldsymbol{\nu}) \\ \mathbf{A}(\boldsymbol{\nu})^{\top} & \boldsymbol{\Sigma}_{\mathbf{H}}(\boldsymbol{\nu})\end{pmatrix}, \tag{S6}$$

in which $\mathbf{A}(\boldsymbol{\nu}) = \text{Cov}\left[\mathbf{H}(\mathbf{Y}_i;\boldsymbol{\nu}), \nabla_{\boldsymbol{\nu}}\log f(\mathbf{Y}_i;\boldsymbol{\nu})\right]$ and $\boldsymbol{\Sigma}_{\mathbf{H}}(\boldsymbol{\nu}) = \text{Cov}[\mathbf{H}(\mathbf{Y}_i;\boldsymbol{\nu})]$.

Now, apply the Delta method to obtain the asymptotic distribution of the reliability coefficient $\varphi(\hat{\boldsymbol{\eta}}(\hat{\boldsymbol{\nu}}))$:

$$\sqrt{n}\left[\varphi(\hat{\boldsymbol{\eta}}(\hat{\boldsymbol{\nu}})) - \varphi(\boldsymbol{\eta}(\boldsymbol{\nu}))\right] = \sqrt{n}\nabla\varphi\left(\boldsymbol{\eta}(\boldsymbol{\nu})\right)^{\top}[\hat{\boldsymbol{\eta}}(\hat{\boldsymbol{\nu}}) - \boldsymbol{\eta}(\boldsymbol{\nu})] + o_p(1)$$
$$\xrightarrow{d} \mathcal{N}\left(0, \nabla\varphi(\boldsymbol{\eta}(\boldsymbol{\nu}))^{\top}\boldsymbol{\Sigma}(\boldsymbol{\nu})\nabla\varphi(\boldsymbol{\eta}(\boldsymbol{\nu}))\right). \tag{S7}$$

Equation S7 is the final expression, identical to Equation 18 in the main document. In the following subsections, we provide the expressions for $\nabla_{\boldsymbol{\nu}}\mathbf{H}(\mathbf{Y}_i;\boldsymbol{\nu})$ and $\nabla\varphi(\boldsymbol{\eta}(\boldsymbol{\nu}))$, derived specifically for PRMSE and CTT reliability of interest.

### A.2 PRMSE

For PRMSE, we defined the function $\mathbf{H}$ in Equation 21 as

$$\mathbf{H}(\mathbf{Y}_i;\boldsymbol{\nu}) = (H_1(\mathbf{Y}_i;\boldsymbol{\nu}), H_2(\mathbf{Y}_i;\boldsymbol{\nu}), H_3(\mathbf{Y}_i;\boldsymbol{\nu}))^{\top}, \tag{S8}$$

in which $H_1(\mathbf{Y}_i;\boldsymbol{\nu}) = \mathbb{E}(\Theta_i|\mathbf{Y}_i;\boldsymbol{\nu})$, $H_2(\mathbf{Y}_i;\boldsymbol{\nu}) = \mathbb{E}(\Theta_i|\mathbf{Y}_i;\boldsymbol{\nu})^2$, and $H_3(\mathbf{Y}_i;\boldsymbol{\nu}) = \text{Var}(\Theta_i|\mathbf{Y}_i;\boldsymbol{\nu})$. The gradients of functions $H_1$, $H_2$, and $H_3$ with respect to $\boldsymbol{\nu}$ are derived as follows:

$$\nabla_{\boldsymbol{\nu}} H_1(\mathbf{Y}_i; \boldsymbol{\nu}) = \nabla_{\boldsymbol{\nu}} \mathbb{E}(\Theta_i | \mathbf{Y}_i; \boldsymbol{\nu}) = \nabla_{\boldsymbol{\nu}} \int \theta_i f(\theta_i | \mathbf{y}_i; \boldsymbol{\nu}) d\theta_i$$
$$= \int \theta_i \phi(\theta_i) \nabla_{\boldsymbol{\nu}} \left[ \frac{f(\mathbf{y}_i | \theta_i; \boldsymbol{\nu})}{f(\mathbf{y}_i; \boldsymbol{\nu})} \right] d\theta_i, \tag{S9}$$

$$\nabla_{\boldsymbol{\nu}} H_2(\mathbf{Y}_i; \boldsymbol{\nu}) = \nabla_{\boldsymbol{\nu}} \mathbb{E}(\Theta_i | \mathbf{Y}_i; \boldsymbol{\nu})^2 = 2 H_1(\mathbf{Y}_i; \boldsymbol{\nu}) \nabla_{\boldsymbol{\nu}} H_1(\mathbf{Y}_i; \boldsymbol{\nu}), \tag{S10}$$

and
$$\nabla_{\boldsymbol{\nu}} H_3(\mathbf{Y}_i; \boldsymbol{\nu}) = \nabla_{\boldsymbol{\nu}} \text{Var}(\Theta_i | \mathbf{Y}_i; \boldsymbol{\nu}) = \nabla_{\boldsymbol{\nu}} \mathbb{E}(\Theta_i^2 | \mathbf{Y}_i; \boldsymbol{\nu}) - \nabla_{\boldsymbol{\nu}} \mathbb{E}(\Theta_i | \mathbf{Y}_i; \boldsymbol{\nu})^2$$
$$= \int \theta_i^2 \phi(\theta_i) \nabla_{\boldsymbol{\nu}} \left[ \frac{f(\mathbf{y}_i | \theta_i; \boldsymbol{\nu})}{f(\mathbf{y}_i; \boldsymbol{\nu})} \right] d\theta_i - \nabla_{\boldsymbol{\nu}} H_2(\mathbf{Y}_i; \boldsymbol{\nu}), \tag{S11}$$

in which the integrals are approximated using quadrature. In Equations S9–S11, the gradient of the bracketed term can be rewritten as:

$$\nabla_{\boldsymbol{\nu}} \left[ \frac{f(\mathbf{y}_i | \theta_i; \boldsymbol{\nu})}{f(\mathbf{y}_i; \boldsymbol{\nu})} \right]$$
$$= \frac{f(\mathbf{y}_i | \theta_i; \boldsymbol{\nu})}{f(\mathbf{y}_i; \boldsymbol{\nu})} \left[ \nabla_{\boldsymbol{\nu}} \log f(\mathbf{y}_i | \theta_i; \boldsymbol{\nu}) - \nabla_{\boldsymbol{\nu}} \log f(\mathbf{y}_i; \boldsymbol{\nu}) \right] \tag{S12}$$
$$= \frac{f(\mathbf{y}_i | \theta_i; \boldsymbol{\nu})}{f(\mathbf{y}_i; \boldsymbol{\nu})} \left[ \nabla_{\boldsymbol{\nu}} \log f(\mathbf{y}_i | \theta_i; \boldsymbol{\nu}) - \frac{\int \nabla_{\boldsymbol{\nu}} \log f(\mathbf{y}_i | \theta_i; \boldsymbol{\nu}) f(\mathbf{y}_i | \theta_i; \boldsymbol{\nu}) \phi(\theta_i) d\theta_i}{f(\mathbf{y}_i; \boldsymbol{\nu})} \right].$$

Recall that $\log f(\mathbf{y}_i | \theta_i; \boldsymbol{\nu}) = \sum_{j=1}^{m} \log f_j(y_{ij} | \theta_i; a_j, c_j)$. For each item $j$, we express the derivative of the log item response function $\log f_j$ with respect to $a_j$ and $c_j$ as follows:

$$\frac{\partial}{\partial a_j} \log f_j(y_{ij} | \theta_i; a_j, c_j) = \theta_i \left[ y_{ij} - f_j(1 | \theta_i; a_j, c_j) \right], \tag{S13}$$

and
$$\frac{\partial}{\partial c_j} \log f_j(y_{ij} | \theta_i; a_j, c_j) = y_{ij} - f_j(1 | \theta_i; a_j, c_j). \tag{S14}$$

To compute PRMSE for the latent variable, we apply the transformation function $\varphi_{\text{PRMSE}}(\mathbf{x}) = (x_2 - x_1^2)/(x_2 - x_1^2 + x_3)$ where $\mathbf{x} = (x_1, x_2, x_3)^\top$. The gradient of $\varphi_{\text{PRMSE}}$ is

$$\nabla \varphi_{\text{PRMSE}}(\mathbf{x}) = \left( \frac{-2x_1 x_3}{(x_2 - x_1^2 + x_3)^2}, \frac{x_3}{(x_2 - x_1^2 + x_3)^2}, \frac{x_1^2 - x_2}{(x_2 - x_1^2 + x_3)^2} \right)^\top. \tag{S15}$$

### A.3 CTT Reliability

For CTT reliability, we defined $\mathbf{H}$ in Equation 29 as

$$\mathbf{H}(\mathbf{Y}_i; \boldsymbol{\nu}) = (H_1(\mathbf{Y}_i; \boldsymbol{\nu}), H_2(\mathbf{Y}_i; \boldsymbol{\nu}), H_3(\mathbf{Y}_i; \boldsymbol{\nu}), \ldots, H_{2+Q}(\mathbf{Y}_i; \boldsymbol{\nu}))^\top, \tag{S16}$$

in which $H_1(\mathbf{Y}_i; \boldsymbol{\nu})$ and $H_2(\mathbf{Y}_i; \boldsymbol{\nu})$ are the same as in PRMSE, and the remaining functions are defined as

$$H_{2+q}(\mathbf{Y}_i; \boldsymbol{\nu}) = H_1(\mathbf{Y}_i; \boldsymbol{\nu}) \frac{f(\mathbf{Y}_i|\theta_{iq}; \boldsymbol{\nu})}{f(\mathbf{Y}_i; \boldsymbol{\nu})}, \quad q = 1, \ldots, Q. \tag{S17}$$

The gradient of Equation S17 is expressed by

$$\begin{aligned}\nabla_{\boldsymbol{\nu}} H_{2+q}(\mathbf{Y}_i; \boldsymbol{\nu}) &= \nabla_{\boldsymbol{\nu}} \left[ H_1(\mathbf{Y}_i; \boldsymbol{\nu}) \frac{f(\mathbf{Y}_i|\theta_{iq}; \boldsymbol{\nu})}{f(\mathbf{Y}_i; \boldsymbol{\nu})} \right] \\ &= \nabla_{\boldsymbol{\nu}} H_1(\mathbf{Y}_i; \boldsymbol{\nu}) \frac{f(\mathbf{Y}_i|\theta_{iq}; \boldsymbol{\nu})}{f(\mathbf{Y}_i; \boldsymbol{\nu})} + H_1(\mathbf{Y}_i; \boldsymbol{\nu}) \nabla_{\boldsymbol{\nu}} \left[ \frac{f(\mathbf{Y}_i|\theta_{iq}; \boldsymbol{\nu})}{f(\mathbf{Y}_i; \boldsymbol{\nu})} \right]. \end{aligned} \tag{S18}$$

In Equation S18, $\nabla_{\boldsymbol{\nu}} H_1(\mathbf{Y}_i; \boldsymbol{\nu})$ is given by Equation S9, and $\nabla_{\boldsymbol{\nu}}[f(\mathbf{Y}_i|\theta_{iq}; \boldsymbol{\nu})/f(\mathbf{Y}_i; \boldsymbol{\nu})]$ is given by S12.

To compute CTT reliability, we apply the transformation function $\varphi_{\text{Rel}}(\mathbf{x}) = (S - x_1^2)/(x_2 - x_1^2)$ where $\mathbf{x} = (x_1, \ldots, x_{2+Q})^\top$, and $S = \sum_{q=1}^{Q} x_{2+q}^2 w_q$. The gradient of $\varphi_{\text{Rel}}$ is

$$\nabla \varphi_{\text{Rel}}(\mathbf{x}) = \left( \frac{2x_1(S - x_2)}{(x_2 - x_1^2)^2}, \frac{-(S - x_1^2)}{(x_2 - x_1^2)^2}, \frac{2w_1 x_3}{x_2 - x_1^2}, \frac{2w_2 x_4}{x_2 - x_1^2}, \ldots, \frac{2w_Q x_{2+Q}}{x_2 - x_1^2} \right)^\top. \tag{S19}$$



\end{document}